\documentclass{aa}

\usepackage{graphicx}
\usepackage{natbib}
\usepackage[normalem]{ulem}
\bibpunct{(}{)}{;}{a}{}{,}
\usepackage{txfonts}
\titlerunning{A parametric study on the formation of ECs and UCDs}
\authorrunning{R.C. Br\"uns et al.}

\begin{document}

\title{A parametric study on the formation of extended star clusters and ultra-compact dwarf galaxies}

\author{R.C. Br\"uns\inst{1} \and P. Kroupa\inst{1} \and M. Fellhauer\inst{2} \and M. Metz\inst{3} \and P. Assmann\inst{2}}

\institute{Argelander-Institut f\"ur Astronomie, Universit\"at Bonn,
            Auf dem H\"ugel 71, D-53121 Bonn, Germany\\
              \email{rcbruens@astro.uni-bonn.de, pavel@astro.uni-bonn.de}
         \and
            Departamento de Astronom\'ia, Universidad de Concepci\'on, Casilla 160-C, Concepci\'on, Chile\\
              \email{mfellhauer@astro-udec.cl, passmann@astro-udec.cl}
	 \and
            Deutsches Zentrum f\"ur Luft- und Raumfahrt, K\"onigswinterer Str. 522-524, D-53227 Bonn, Germany\\
              \email{manuel.metz@dlr.de}     
             }
	     
\abstract
   {In the last decade, very extended old stellar clusters with masses in the range from
  a few $10^{4}$ to $10^{8}$ M$_{\sun}$ have been found in various types of 
galaxies in different environments. Objects with masses comparable
to normal globular clusters (GCs) are called extended clusters (ECs), while objects with
masses in the dwarf galaxy regime are called ultra-compact dwarf galaxies (UCDs).
In heavily interacting galaxies star clusters tend to form in larger conglomerations called 
star cluster complexes (CCs). The individual star clusters in a CC can merge 
and form a variety of spheroidal stellar objects.}
   {The parametric study aims to analyze how the structural parameters of the final 
merger objects correlate with the underlying CC parameter space.}
   {In this work we systematically scan a suitable parameter space for CCs and perform numerical 
simulations to study their further fate. The varied sizes and masses of the CCs cover a 
matrix of 5x6 values with CC Plummer radii between 10 - 160 pc and CC masses between 
$10^{5.5}$ - $10^{8}$ M$_{\sun}$, which are consistent with observed CC parameters. 
The CCs of the parametric study are on orbits with galactocentric distances between 
20 kpc and 60 kpc. In addition, we studied also the evolution of CCs on a circular orbit 
at a galactocentric distance of 60 kpc to verify that also extremely extended ECs and UCDs 
can be explained by our formation scenario.}
   {All 54 simulations end up with stable merger objects, wherein 26 to 97\% of the initial CC mass
is bound. The objects show a general trend of increasing effective radii with increasing mass. 
Despite the large range of input Plummer radii of the CCs (10 to 160 pc) the effective radii of 
the merger objects are constrained to values between 10 and 20 pc at the low mass end and to 
values between 15 and 55 pc at the high mass end. The structural parameters of the models are 
comparable to those of the observed ECs and UCDs. The results of the circular orbits demonstrate 
that even very extended objects like the M31 ECs found by Huxor in 2005 and the very extended 
($r_{\rm eff} >$ 80 pc), high-mass UCDs can be explained by merged cluster complexes in regions 
with low gravitational fields at large galactocentric radii.}
   {We conclude that the observed ECs and UCDs can be well explained as evolved star cluster 
   complexes.}
   
	       
  \maketitle

  \section{Introduction}\label{introduction}

  Globular clusters (GCs) are very old stellar objects with typical masses between 
$10^4$ M$_{\sun}$ and $10^6$ M$_{\sun}$ (corresponding roughly to total luminosities between 
$M_V = -5$ to $M_V = -10$), having in general compact sizes with half-light radii of a few pc. 
This morphology makes them easily observable also in external galaxies with modern telescopes 
\citep[see][and references therein]{brodie06}.

The \object{Milky Way} has a rich GC system containing 150 GCs \citep{harris}. Most of them are compact
with sizes of a few pc. Only 13 GCs (or 9\%) have an effective radius larger than 10~pc. 
Most of these extended clusters (ECs) are fainter than about $M_{\rm V} = -7$, only \object{NGC\,2419}, 
having a half-light radius of about 20~pc, has a high luminosity of about 
$M_{\rm V} = -9.4$ mag. Further ECs in the vicinity of the Milky Way have been found in the 
\object{LMC} and the \object{Fornax dwarf galaxy} \citep{mackey04,vandenbergh04,mvdm}.

Comparable objects have also been detected around other galaxies. 
\citet{huxor04} found three ECs around \object{M31}, which have very large radii above 30~pc. 
These clusters were detected by chance as the automatic detection algorithms of the 
MegaCam Survey discarded such extended objects as likely background contaminations. 
Follow-up observations by \citet{mackey06}, using the ACS camera of the Hubble Space Telescope 
(HST), resolved the ECs into stars proving their nature as M31 clusters. They also detected a 
fourth EC around M31. The M31 ECs have masses of the order of $10^5$ M$_{\sun}$. Further 
observations increased the number of ECs in M31 to 13 \citep{huxor08}. 
However, \cite{huxor11} showed that the previous estimates of the effective radii were
considerably too large. The new size estimate are well below 30~pc. 
\citet{chandar04} observed a part of the disks of the nearby galaxies \object{M81}, \object{M83}, 
\object{NGC6946}, \object{M101}, and \object{M51} using HST and found ECs with effective radii 
larger than 10 pc in four of them. M51 showed a very high fraction of ECs in the observed area: 
8 of 34 GCs (24\%). ECs are now detected in all types of galaxies from dwarfs to 
ellipticals \citep[e.g.][]{larbro00,harris02,lee05,peng,chisa,stonkute,georgiev,dacosta}.

\citet{hilker99} and \citet{drinkwater00} discovered in the \object{Fornax Cluster} compact objects 
with luminosities above the brightest known GCs and which were not resolved by ground-based 
observations. These objects have masses between a few $10^{6}$~M$_{\sun}$ and $10^{8}$~M$_{\sun}$ 
and effective radii between $r_{\rm eff} = 10$ and 100~pc. 
\citet{drinkwater00} interpreted these objects as a new type of galaxy and reflected this 
interpretation in the name ``ultra-compact dwarf galaxy'' (UCD). \citet{bekki01} suggested 
that UCDs are the remnants of dwarf galaxies which lost their dark matter halo and 
all stars except their nucleus. 
Next to the interpretation as a galaxy, UCDs were also considered as high-mass versions 
of normal GCs \citep{mieske02}, or as merged massive complexes of star clusters 
\citep{krou98,fellhauer02a}.                                                                                                             

Many UCDs have been found now. Next to the Fornax Cluster, they have been observed in the 
galaxy cluster \object{Abell~1689} \citep{mieske04}, around \object{M87} in the \object{Virgo Cluster} 
\citep{hasegan,evstigneeva07}, the \object{Centaurus Cluster} \citep{mieske07}, 
the \object{Coma Cluster} \citep{madrid}, and \object{Abell S0740} \citep{blakeslee}. 
While most known UCDs belong to galaxy clusters, they have also been observed in rather isolated      
environments, e.g. in the Sombrero galaxy (\object{M104}) by \citet{hau}.                                   
                            
\citet{forbes08} and \citet{mieske08} analyzed larger samples of UCDs. They find that normal 
and extended star clusters and UCDs form a coherent data set where size and mass-to-light ratio increase
continuously with their total mass and concluded that UCDs are more likely bright extended 
clusters than naked cores of stripped dwarf galaxies. The marginally enhanced 
mass-to-light ratios of UCDs can be explained by slightly modified initial stellar mass functions 
\citep{miekrou08,dabringhausen09}.   

High-resolution HST imaging of gas-rich galaxies experiencing major interactions 
has resolved very intense star formation bursts. The bursting regions are typically
located within the severely perturbed disks or tidal tails and are constrained to
small complexes that contain a few to hundreds of young massive star clusters. Examples of such 
systems are the knots in the \object{Antennae galaxies} \citep{whitmore95, whitmore99}, 
the complexes in the \object{NGC\,7673} star-burst \citep{homeier}, \object{M82} \citep{konstantopoulos}, 
\object{Arp24} \citep{cao}, the ``bird's head galaxy'' \object{NGC\,6745} \citep{de_grijs03}, 
\object{NGC\,6946} \citep{larsen02}, \object{Stephan's Quintet} \citep{gallagher01}, and \object{NGC\,922} 
\citep{pellerin}.

The masses of such complexes vary from about $10^6$ M$_{\sun}$ up to a few~$10^8$ M$_{\sun}$.
\citet{bastian06} observed star cluster complexes (CCs) in the Antennae with masses of the order 
$\approx 10^{6}$~M$_{\sun}$ and diameters of the order 100 to 200 pc. 
\citet{pellerin} found young massive CCs with masses between $10^6$ M$_{\sun}$ and $10^{7.5}$ M$_{\sun}$ 
and diameters between 600 pc and 1200 pc in the collisional ring galaxy NGC\,922. 
One of the most extended CCs has been observed by \cite{tran} in the tail of the ``Tadpole galaxy'' 
\object{UGC\,10214}. This CC, which has a mass of the order $10^6$ M$_{\sun}$, has an effective radius of
160 pc and a diameter of about 1500 pc.

\citet{mengel08} observed individual young ($\approx$10 Myr) star clusters associated with 
CCs in the Antennae and \object{NGC\,1487}. They compared dynamical mass estimates with 
derived photometric masses and found them in excellent agreement, implying that most of them 
survived the gas removal phase and are bound stellar objects. These young clusters are 
sufficiently stable to be used as building blocks for numerical simulations.
\citet{bastian09} found three 200 to 500 Myr old, apparently stable clusters in the Antennae with 
very high radial velocities relative to the galactic disk, indicating that these star clusters 
will most likely become future halo objects. One cluster is surrounded by so far unmerged stellar 
features in its vicinity. 

Since galaxy-galaxy mergers are anticipated to have been more common during early cosmological times 
it is expected that star formation in CCs has been a significant star formation mode during this epoch. 
Indeed, the preponderance of clumpy galaxies \citep[and references therein]{elmegreen07} indicates
that early gas-rich galaxies went through an epoch of profuse CC formation. 
We propose a formation scenario for ECs and UCDs where massive 
complexes of star clusters were formed during a heavy interaction between gas-rich galaxies, which 
lead to the formation of various kinds of objects via merging of its constituent star clusters.
It has already been shown in previous papers that CCs can merge to form a variety of spheroidal 
stellar objects, such as UCDs, ECs, faint fuzzies and possibly dwarf spheroidal galaxies 
\citep{krou98,fellhauer02a,fellhauer02b,bekki04,bruens09, bruens10}. In particular, the young UCD \object{W3} 
is most naturally understood to be a merged massive CC \citep{fellhauer05}. 

In the present paper, we broaden the scope to analyze how the structural parameters of the final 
merger objects correlate with the underlying CC parameter space. We systematically scan a suitable 
parameter space for CCs, covering a large range of CC masses and sizes. 
In Section \ref{observations} we summarize the status quo of available observational data on 
ECs and UCDs used as input for this parametric study. In Section \ref{simulations}, we describe 
the method and the parameters used for the calculations. The results are presented in 
Section \ref{results} and discussed in Section \ref{discussion}.

\section{Observations} \label{observations}

\begin{figure}
\centering
\includegraphics[width=8.9cm]{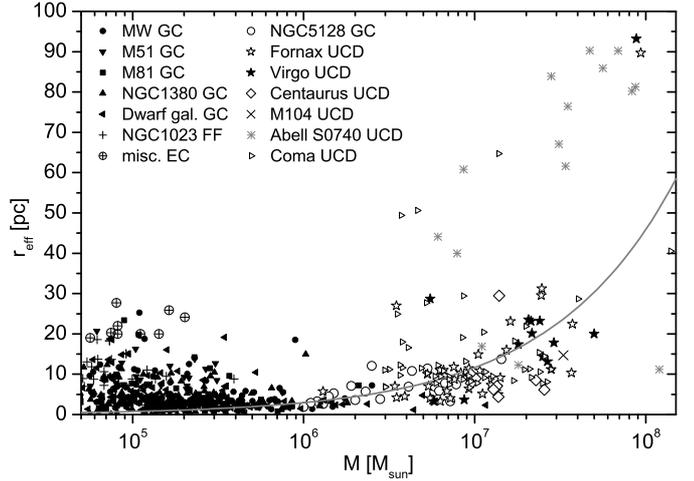}
\caption{Overview of observed effective radii and mass estimates of GCs, ECs and UCDs 
(see Sect. \ref{observations} for details). The grey curve indicates the general trend
of increasing $r_{\rm eff}$ as parameterized by \cite{dabringhausen08}.}
\label{fig_ucdgc}
\end{figure}

As already indicated in the previous section, GC-like objects with effective radii above 10 pc have been 
found in various environments covering a large mass range. To allow for an overview of their
parameters, we compiled a list of effective radii and mass estimates of GCs, ECs, and UCDs. 
The respective publications used to compile the data-base are summarized in the following subsections.

\subsection{GC systems and ECs}
For the Milky Way and its satellites, we use the catalog of resolved massive star clusters of \cite{mvdm}.

The Milky Way data were complemented by the GC tables from \cite{chandar04} for the spiral galaxies 
M51 and M81. In addition, the catalog of GCs in the lenticular galaxy \object{NGC\,1380} \citep{chisa} and
the faint fuzzy star clusters found in the lenticular galaxy \object{NGC\,1023} by \cite{larbro00} were used.
The catalog of GCs in 68 dwarf galaxies \citep{georgiev} was added to account for GCs and ECs found
in dwarf galaxies.

Next to the aforementioned GC catalogs miscellaneous ECs from M31 \citep{huxor11}, M33 \citep{stonkute}, 
\object{Scl-dE1} \citep{dacosta}, and \object{NGC\,6822} \citep{lee05} were added to the sample.

\cite{mvdm} estimated cluster masses for all objects using population-synthesis models to define a V-band 
mass-to-light ratio for every cluster. We use a typical mass-to-light ratio of $M/L_{V} = 2$, 
which is consistent with the results of \cite{mvdm}, to estimate masses for the remaining objects
of this section.

\subsection{UCDs and massive GCs} 
\cite{rejkuba} and \cite{taylor} analyzed a sample of massive GCs in the nearby giant elliptical 
galaxy \object{NGC\,5128} (Centaurus A). Their sample contains 22 massive GCs with masses larger than 
$10^{6}$ M$_{\sun}$. 

The parameters of UCDs were compiled for the Fornax Cluster \citep{mieske08,richtler,evstigneeva08}, 
the Virgo Cluster \citep{hasegan,evstigneeva07,evstigneeva08}, the Centaurus Cluster \citep{mieske07},
and the Coma Cluster \citep{madrid}. 
In addition, the UCD found in the Sombrero galaxy (M104) by \cite{hau} was considered.

\cite{blakeslee} identified 15 UCD candidates in the ABELL S0740 cluster, which have rather large effective 
radii. We added these UCD candidates to our list, but it should be noted that these UCDs are not yet confirmed
members of the ABELL S0740 cluster and might therefore be background objects.

A large fraction of UCDs has reliable mass estimates. We used a typical mass-to-light ratio for 
UCDs of  $M/L_{V} = 4$ \citep{mieske08} to derive mass estimates for those UCDs, where no mass 
estimate was published so far.

\subsection{Overview of GC, EC, and UCD parameters}

Figure~\ref{fig_ucdgc} shows the effective radii of the GCs, ECs, and UCDs as a function of their 
estimated masses. Below masses of $10^{6}$ M$_{\sun}$, the vast majority of clusters have effective radii
of a few pc. Nevertheless, a few dozens of objects have effective radii larger than 10 pc. Most of them 
have masses of the order $10^{5}$ M$_{\sun}$. In contrast, only very few ECs are found at masses of the 
order $10^{6}$ M$_{\sun}$.

For masses above $10^{6}$ M$_{\sun}$, there is no clear concentration at low effective radii. 
The radii are more evenly distributed with a clear trend of increasing $r_{\rm eff}$ with increasing
masses. Above $10^{7.5}$ M$_{\sun}$ all objects have effective radii above 10 pc. The general trend
of increasing $r_{\rm eff}$, which is added as a grey curve in Fig. \ref{fig_ucdgc}, was parameterized 
by \cite{dabringhausen08}. While this line provides a trend for the majority of massive GCs and UCDs,
the scatter is quite large and a number of objects are located far from this line. 

Most UCDs show $r_{\rm eff}$ smaller than 35 pc. However, one UCD in the Fornax Cluster 
and one in the Virgo Cluster have effective radii larger than 90 pc. These two UCDs show a core-halo
structure, where the cores have an effective radius of about 10 pc \citep{evstigneeva07}. 
In addition, four UCD candidates in the Coma cluster \citep{madrid} and the majority of the
UCD candidates in Abell S0740 \citep{blakeslee} are considerably larger than 35 pc. 
As most of these objects are not yet confirmed as cluster members, they might also be background galaxies.

\begin{figure}
\centering
\includegraphics[width=8.9cm]{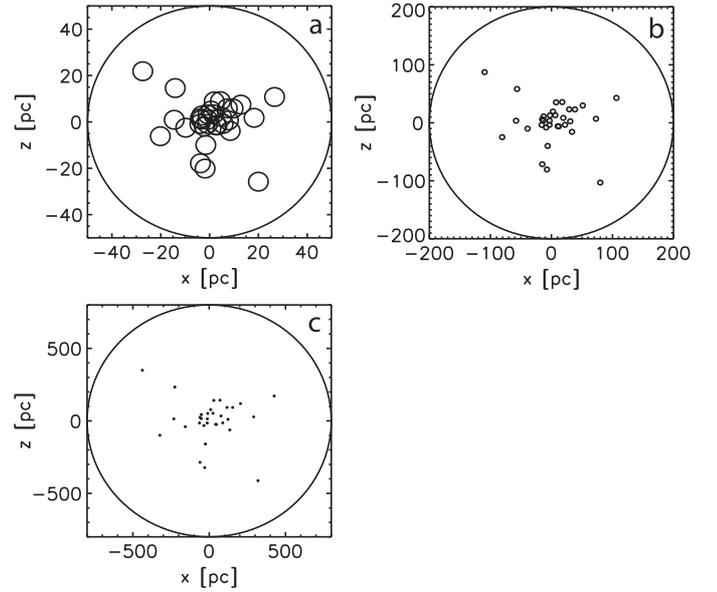}
\caption{Three exemplary initial distributions of star clusters (small circles with radius $R_{\rm
pl}^{\rm SC}$) in a CC (surrounding circle with radius $R_{\rm cut}^{\rm CC}$) projected onto the 
xz-plane. \textbf{(a)} $R_{\rm pl}^{\rm CC} = 10$ pc, \textbf{(b)} $R_{\rm pl}^{\rm CC} = 40$ pc, and 
\textbf{(c)} $R_{\rm pl}^{\rm CC} = 160$ pc. They have $\alpha$-values of 0.4, 0.1, and 0.025, respectively.}
\label{fig_ini_configs}
\end{figure}

\begin{figure}
\centering
\includegraphics[width=8.9cm]{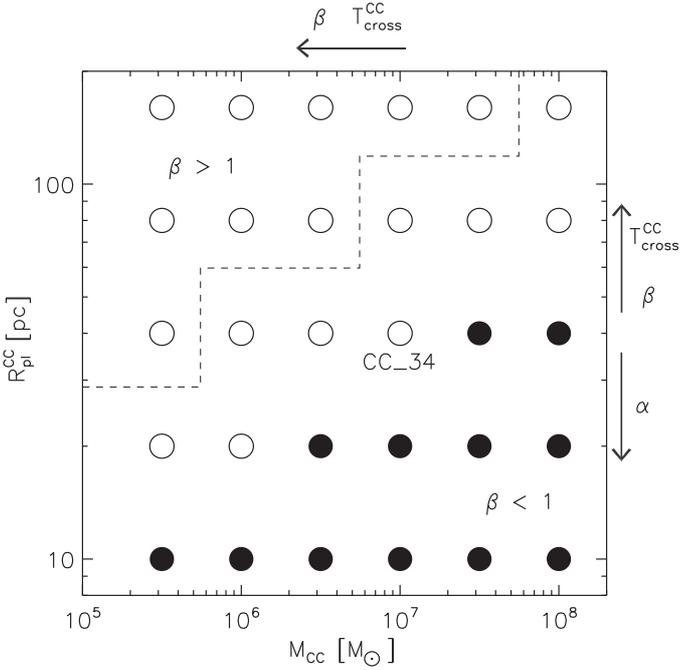}
\caption{Parameter space for the CC models. The parameters CC size, $R_{\rm pl}^{\rm CC}$, and CC mass, 
$M^{\rm CC}$, constitute a matrix of 5x6 values. All circles mark the 30 simulations on orbit 1. The 
open circles indicate the 18 additional simulations on orbit 2. The arrows indicate the increase of 
CC-crossing time $T_{\rm cross}^{\rm CC}$, and $\alpha$- and $\beta$-values (see Sect. \ref{alpha_beta}). 
The dashed line separates the CC models with $\beta < 1$ from the CC models with $\beta > 1$. CC\_34 is 
an example of the nomenclature of the models. The first index is the row number which corresponds to a 
CC Plummer radius $R_{\rm pl}^{\rm CC}$. The second index indicates the column synonymous to a CC mass $M^{\rm CC}$.}
\label{fig_grid}
\end{figure}

\section{Numerical method and set-up} \label{simulations}

\newcommand{\cpp}{C{\footnotesize\raise.4ex\hbox{+\kern-.2em+}}}
\newcommand{\sbpp}{{\scshape Superbox{\footnotesize\raise.4ex\hbox{+\kern-.2em+}}}}
\newcommand{\subo}{{\scshape Superbox}}

\subsection{Scenario and model parameters} \label{model_params}

The formation scenario described in this paper starts with newly born complexes of star clusters
covering a large range in masses and sizes. We model the dynamical evolution of various CCs 
leading to merger objects. We do not, however, consider the galaxy-galaxy interaction, which 
formed the CCs in the first place. 

Individual young massive star clusters were analyzed in detail by \citet{mengel08} 
and \citet{bastian09}, leading to a combined sample of 25 objects. The median effective radius
of these 25 young massive star clusters is 4 pc. We use this value for our individual star clusters,
which are the building blocks of the CC models. 

All CC models in this paper consist of $N_{\rm 0}^{\rm CC}$ = 32 star clusters.
The individual star clusters are represented by Plummer spheres \citep{plum1911, krou08} 
with $N_{\rm 0}^{\rm SC}$ = 100\,000 particles. The Plummer radius, which corresponds to 
the effective radius, is chosen to be 4 pc for all models. We select a cutoff radius of 
$R_{\rm cut}^{\rm SC}= 5 R_{\rm pl}^{\rm SC} = 20$ pc. For each CC, the 32 star clusters have 
the same mass, which is 1/32 of the corresponding CC mass. The initial velocity distribution 
of the star clusters is chosen such that they are in virial equilibrium.

The observed CCs show a clear concentration of star clusters towards their centers 
\citep{tran,bastian06,pellerin}. Unfortunately, no detailed observational constraints 
on the distribution of star clusters in the complex and their dynamical state are available. 
In the absence of observed density profiles of CCs, we choose a simple model and distribute
the star clusters in the CC models according to a Plummer distribution, which is truncated at 
the cutoff radius $R_{\rm cut}^{\rm CC} = 5 R_{\rm pl}^{\rm CC}$. 
This cutoff radius is large enough to prevent a clear break or edge in the spatial distribution and 
small enough to avoid single star clusters at very large distances that would be stripped away immediately. 
If we would increase our cutoff radius $R_{\rm cut}^{\rm CC}$ from 5 times the Plummer radius 
$R_{\rm pl}^{\rm CC}$ to infinity, we would have only one or two star clusters beyond our actual 
cutoff radius. Hence, the exact value of the cutoff radius will have a negligible impact on the results.
Figure \ref{fig_ini_configs} shows three exemplary initial distributions of star clusters in the 
CC. We used the same seeds for the random number generator to generate the same distribution of 
star clusters scaled to the corresponding Plummer radius of the CC model. 
In a previous paper on the extended Milky Way cluster NGC\,2419 \citep{bruens10}, we have demonstrated 
that the exact initial distribution of star clusters in an extended CC leads to variations in
the structural parameters mass and size of the order 10 to 20\%. 
The initial velocity distribution of the CC models is chosen such that 
each CC is in virial equilibrium. A detailed description of the generation of initial coordinates 
(space and velocity) for Plummer models is given in the appendix of \cite{aarseth}.

The model parameters constitute a matrix of 5x6 values (Fig. \ref{fig_grid}) with CC Plummer 
radii of $R_{\rm pl}^{\rm CC} = 10, 20, 40, 80, 160$ pc and total CC masses of 
$M^{\rm CC} = 10^{5.5}, 10^{6}, 10^{6.5}, 10^{7}, 10^{7.5}, 10^{8}$ M$_{\sun}$. 
The range of sizes and masses are motivated by the observed parameters of ECs and UCDs 
(see Sect. \ref{observations}) and they are consistent with observations of CCs 
(see Sect. \ref{introduction}). For the CC masses $M^{\rm CC} = 10^{6.5}$ and $10^{7.5}$ M$_{\sun}$
two additional models with $R_{\rm pl}^{\rm CC} =$ 240 and 360 pc were considered. 

\begin{figure}
\centering
\includegraphics[width=8.9cm]{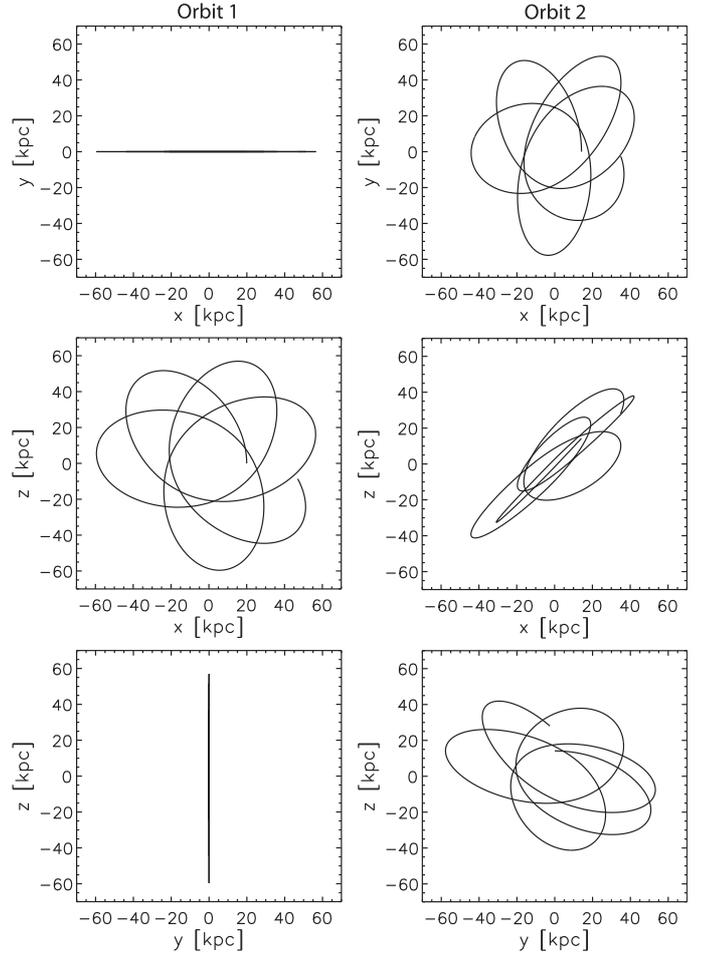}
\caption{The orbits projected to the xy-, the xz-, and the yz-plane. Orbit 1 is a polar orbit and orbit 2 is
an inclined orbit between galactic radii 20 and 60 kpc. }
\label{fig_orbits}
\end{figure}

\subsection{Gravitational potential and orbits} \label{potential_mw}

ECs are found near late type disk galaxies, lenticular galaxies, elliptical, and dwarf galaxies, 
while UCDs are predominantly found close to giant elliptical galaxies. As the gravitational potential has a 
larger impact on the low-mass objects and as we like to use the same potential for all computations, 
we chose an analytical Milky-Way-like potential consisting of a disk-, a bulge-, and a halo component. 
The coordinate system is chosen such that the disk of the host galaxy lies in the xy-plane.
The disk is modeled by a Miyamoto-Nagai potential \citep{miya1975},
\begin{eqnarray}
\Phi_\mathrm{disk}(R,z)= -\frac{G\,M_\mathrm{d}}{\sqrt{R^{2}+(a_\mathrm{d}+\sqrt{z^{2}+b_\mathrm{d}^{2}})^{2}}}, 
\end{eqnarray}
with $M_{\rm d}=1.0 \times 10^{11}$ M$_{\sun}$, $a_{\rm d}=6.5$ kpc, and $b_{\rm d}=0.26$ kpc. 
The bulge is represented by a Hernquist potential \citep{hern1990},
\begin{eqnarray}
\Phi_\mathrm{bulge}=-\frac{G \, M_\mathrm{b}}{r+a_\mathrm{b}},
\end{eqnarray}
with $M_{\rm b}=3.4 \times 10^{10}$ M$_{\sun}$, and $a_{\rm b}=0.7$ kpc.
The dark matter halo is a logarithmic potential,
\begin{eqnarray}
\Phi_\mathrm{halo}(r)=\frac{1}{2} \, v_\mathrm{0}^{2} \, {\rm ln}(r^{2}+r_\mathrm{halo}^{2}), 
\end{eqnarray}
with $v_{\rm 0}=186.0$ km s$^{-1}$, and $r_{\rm halo}=12.0$ kpc. This set of parameters gives a 
reasonable Milky-Way-like rotation curve.

In this paper, we focus on ECs and UCDs located far from the galactic disk in the halo of the respective 
galaxies. As orbital parameters for such objects are unknown, we chose a polar orbit between galactic 
radii 20 and 60 kpc for our simulations. These values are motivated by the projected distances to
the M31 ECs of 13 to 60 kpc \citep{mackey06} and the projected distances of Fornax UCDs between
8 and 74 kpc \citep{mieske08}.
Figure \ref{fig_orbits} illustrates the chosen orbit, which has an orbital period of about 860 Myr.  

In our formation scenario, the CCs are most likely formed at the peri-galactic passage of the 
parent galaxy where the impact of the interaction is strongest. Therefore, we start our calculations
at the peri-galactic distance and integrate all models up to 5 Gyr. We stop the integrations at
5 Gyr to save computing time, as the structural parameters change only very slightly afterwards.

To analyze the impact of a polar orbit relative to an inclined orbit, we recalculated a subset 
of our models also on an inclined orbit (see Fig. \ref{fig_orbits}, Orbit 2). 
The orbit is expected to have its largest impact on the most extended and lowest mass CC models. 
The additional computations are indicated by open circles in Fig. \ref{fig_grid}. 

In addition, for the most extended and least massive model ($R_{\rm pl}^{\rm CC} =$ 160 pc and 
$M^{\rm CC} = 10^{5.5}$ M$_{\sun}$), where the tidal field has the largest impact, and for the 
most extended and most massive model ($R_{\rm pl}^{\rm CC} =$ 160 pc and $M^{\rm CC} = 10^{8}$ M$_{\sun}$) 
complementary calculations on a circular orbit at a galactocentric distance of 60 kpc were performed. 

\subsection{Numerical method} \label{method}
The numerical modeling was performed with the particle-mesh code \sbpp\ developed by \cite{metz}. 
It is a new C++ implementation of the FORTRAN particle-mesh code \subo\ \citep{fell00} using object oriented 
programming techniques. \sbpp\ makes particular optimal use of modern multi-core processor technologies.
The code solves the Poisson equation on a system of Cartesian grids. 

In order to get good resolution of the star clusters two grids with high and medium resolution 
are focused on each star cluster following their trajectories. The individual high resolution 
grids have a size of $\pm$80 pc and cover an entire star cluster, whereas the medium resolution grid 
of every star cluster has a size between $\pm$800 pc and $\pm$1200 pc embedding the whole initial CC. 
The local universe is covered by a fixed coarse grid with a size of $\pm$70 kpc, which contains the 
orbit of the CC around the center of the galaxy. All grids contain 128$^{3}$ grid cells.

The galaxy is represented by an analytical potential (see Sect. \ref{potential_mw}). 
For each particle in the CC the acceleration from the galactic potential is added as an analytical 
formula to the grid-based acceleration computed by solving the Poisson equation.

\subsection{General parameters governing the merging process} \label{alpha_beta}

\begin{figure}
\centering
\includegraphics[width=8.9cm]{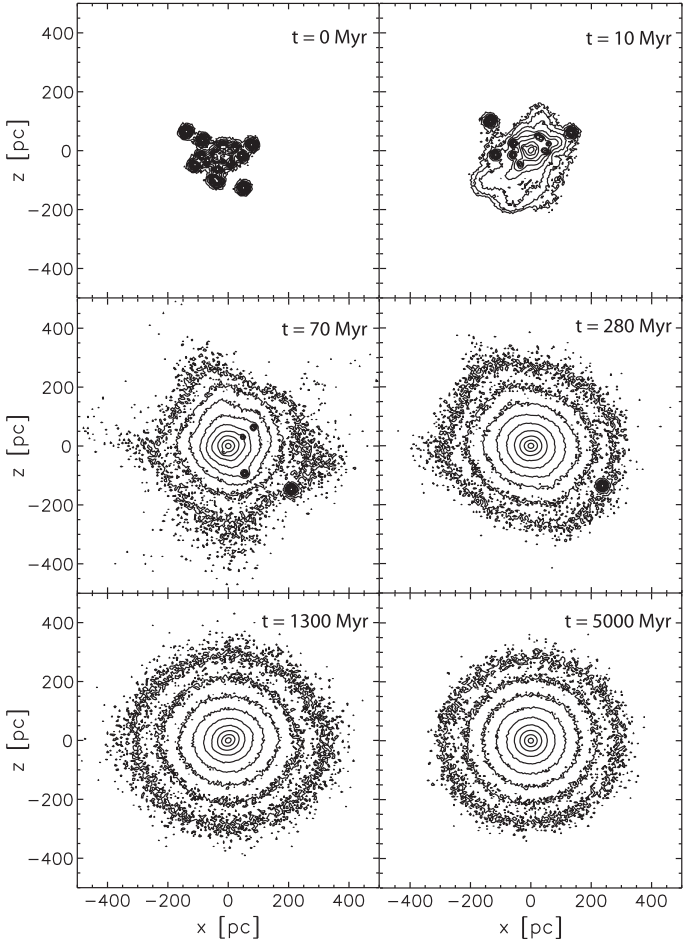}
\caption{Time evolution of the merger object of model CC\_34 ($M^{\rm CC} = 10^{7}$~M$_{\sun}$,
 $R_{\rm pl}^{\rm CC} =$ 40 pc). 
Contour plots on the xz-plane displayed at $t$ = 0, 10, 70, 280, 1300 and 5000 Myr. 
The lowest contour level corresponds to 5 particles per pixel. The pixel size is 5 pc. 
This yields 0.625 M$_{\sun}$ pc$^{-2}$. The contour levels increase further by a factor of 3.}
\label{fig_time_evol}
\end{figure}

The formation process of the merger object depends on the compactness of the initial CC. A measure 
of how densely a CC is filled with star clusters for an equal number $N_{\rm 0}^{\rm CC}$ of star clusters is given
by the parameter $\alpha$ \citep{fell02a},
\begin{eqnarray} \alpha = \frac{R_\mathrm{pl}^\mathrm{SC}}{R_\mathrm{pl}^\mathrm{CC}}, \end{eqnarray} 
where $R_{\rm pl}^{\rm SC}$ and $R_{\rm pl}^{\rm CC}$ are the Plummer radius of a single star cluster and the Plummer 
radius of the CC, respectively. In general high values of $\alpha$ accelerate the merging process because the 
star clusters already overlap in the center of the CC, whereas low values hamper the merging process. Our models cover
$\alpha$-values of 0.4, 0.2, 0.1, 0.05 and 0.025. High values of $\alpha$ ($ \ge 0.1 $) correspond to 
compact CCs with overlapping star clusters in the center (Fig. \ref{fig_ini_configs}a and b) where 
the majority of star clusters merge within a few Myr (Fig. \ref{fig_time_evol}). Low values of 
$\alpha$ ($ \le 0.05 $) correspond to extended CCs (Fig. \ref{fig_ini_configs}c) where 
the merging process can take up to several hundred Myr. The parameter $\alpha$ for the model matrix is 
shown in Fig. \ref{fig_grid} increasing from extended to compact CCs. 

Also the tidal field has to be taken into account as it counteracts the merging process. An estimate of the influence of the 
tidal field on the CC is given by the parameter 
\begin{eqnarray} \beta =\frac{R_\mathrm{cut}^\mathrm{CC}}{r_\mathrm{t}^\mathrm{CC}} \end{eqnarray} \citep{fell02a}, 
which is the ratio of the cutoff radius $R_{\rm cut}^{\rm CC}$ of the CC and 
its tidal radius $r_{\rm t}^{\rm CC}$. An order of magnitude estimate of the tidal radius 
is given by \cite{king62},
\begin{eqnarray}
r_\mathrm{t}^\mathrm{CC} = R_\mathrm{p} \left(\frac{M_\mathrm{CC}}{M_\mathrm{gal,R_p}(3+e)}\right)^{1/3}\label{rtidaleq},
\end{eqnarray}
where $M_{\rm CC}$ is the mass of the CC, $M_{\rm gal,R_p}$ is the galaxy mass within $R_{\rm p}$, 
$R_{\rm p}$ and $R_{\rm a}$ are the peri- and apo-galactic distances, and 
$e=(R_{\rm a}-R_{\rm p})/(R_{\rm a}+R_{\rm p})=0.5$ is the eccentricity of the orbit. 
If the star cluster distribution lies within the tidal radius of the CC ($\beta < 1$) the influence 
of the tidal field on the merging process is small. Almost all star clusters will 
merge and only a few are able to escape by chance. However, in case of $\beta > 1$, a considerable
fraction of star clusters can leave the CC before participating in the merging process. 
The larger the value of $\beta$, the larger the impact of the tidal field on the formation process 
of the merger object. The $\beta$-value distribution of the CC models is displayed in 
Fig. \ref{fig_grid}. The low-mass extended CC models have $\beta$-values $> 1$. 
The $\beta$-values increase with CC size and decrease with CC mass.

For a detailed analysis of merging processes the reader is referred to \cite{fell02a}.

\begin{figure}
\centering
\includegraphics[width=8cm]{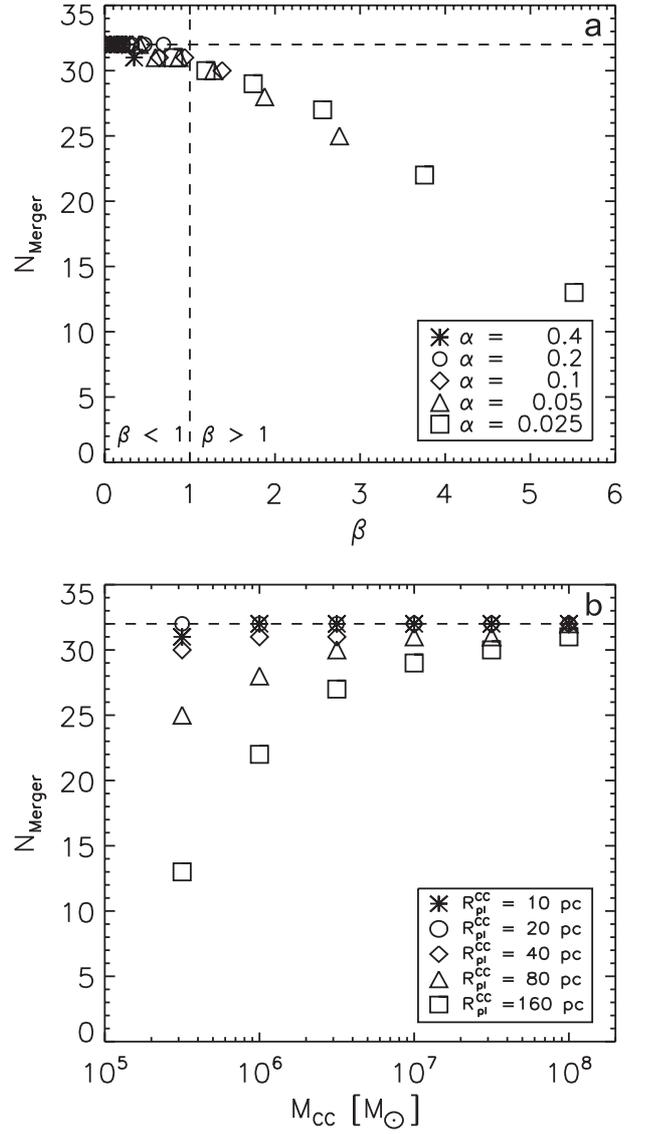}
\caption{\textbf{(a)} Number of merged star clusters versus parameter $\beta$ for different values of $\alpha$
for orbit 1 after 5 Gyr. The horizontal dashed line
marks the number of star clusters $N_{\rm 0}^{\rm CC}$ = 32 in the initial CC. The vertical dashed line
separates the CC models with $\beta < 1$ from the CC models with $\beta > 1$. 
\textbf{(b)} Number of merged star clusters against CC mass for different CC sizes. The horizontal dashed line
denotes the number of star clusters in the initial CC as in Fig. \ref{fig_merger.eps}a.}
\label{fig_merger.eps}
\end{figure}

\section{Results} \label{results}

We carried out 54 different numerical simulations to study the influence of varying initial CC parameters. 

\subsection{Time evolution of the merging process} \label{timeevolution}

To illustrate the merging process, the evolution of model CC\_34 
($R_{\rm pl}^{\rm CC} =$ 40 pc, $M^{\rm CC} = 10^{7}$~M$_{\sun}$) is shown in 
Fig. \ref{fig_time_evol} as contour plots on the xz-plane. 
The snapshots were taken at t = 0, 10, 70, 280, 1300 and 5000 Myr. 
The initial CC has an $\alpha$-value of 0.1. At $t$ = 10 Myr (top panel) the merger object is 
already in the process of formation. In the course of time more and more star clusters are captured by the 
merger object and the merger object becomes more extended. At $t$ = 70 Myr which is about 10 CC-crossing times, 
the vast majority of star clusters has merged into the merger object. Another 20 CC-crossing-times later the 
merging process is almost completely terminated and 31 out of 32 star clusters have merged. There is still a 
close companion star cluster as a satellite of the merger object which eventually (after $t$ = 1300 Myr) also 
falls into the merger object. After the merging process has been completely terminated 
the merger object becomes slightly smaller and reaches a stable state within a few Gyr. As the structural
parameters vary only marginally after a couple of Gyr the simulations are terminated at $t$ = 5 Gyr. 

The general merging process is very similar for all models, but the corresponding time-scale 
varies considerably. A typical timescale for a CC is the crossing time of a star cluster through
the CC, 
\begin{eqnarray}
T_\mathrm{cross}^\mathrm{CC} = \left(\frac{3 \pi}{32}\right)^{-1.5} \sqrt{\frac{(R_\mathrm{pl}^\mathrm{CC})^3}{GM_\mathrm{CC}}}\label{Tcrosseq},
\end{eqnarray}
The values of the crossing time of the CC cover a range between 0.3 Myr (CC\_16) and 340 Myr (CC\_51).
The general trend of the crossing time is indicated by arrows in Fig.~\ref{fig_grid}. 
The results of our calculations show that on average half of the individual star clusters have merged 
after approximately two-and-a-half $T_{\rm cross}^{\rm CC}$.

\subsection{Number of merged star clusters} \label{number_of_merger}

The larger the impact of the tidal field, i.e. the larger the value of $\beta$, the smaller is the 
number of merging events. Fig.~\ref{fig_merger.eps}a shows the number of merged star clusters as 
a function of $\beta$. For models with $\beta < 1$, practically all star clusters merge, while for 
larger values of $\beta$ an increasing number of star clusters are able 
to escape and align along the orbit. 

Figure~\ref{fig_merger.eps}b demonstrates how the number of merged star clusters depends on the 
mass and size of the initial CC. The number of merged star clusters increases with CC mass and 
decreases with CC size. The number of merged clusters becomes as low as 13 for the least massive and most
extended model ($M^{\rm CC} = 10^{5.5}$ M$_{\sun}$ and $R_{\rm pl}^{\rm CC} = 160$ pc).
As the unmerged clusters remain compact GCs, the merging of 13 clusters to one merger objects leads 
to 19 compact GCs originating from the same initial CC.

\begin{figure}
\centering
\includegraphics[width=8cm]{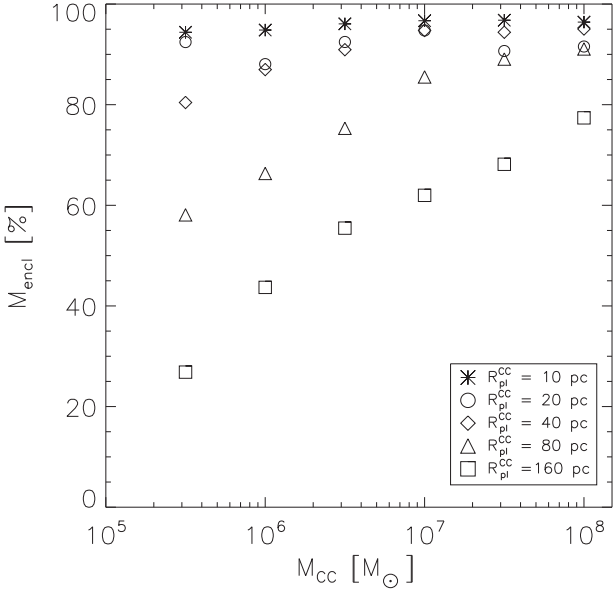}
\caption{Remaining fraction of mass of the initial CC which ended up in the merger object vs. the 
initial CC mass for orbit 1 after 5 Gyr.}
\label{fig_M_encl_M_CC.eps}
\end{figure}

\begin{figure}
\centering
\includegraphics[width=8cm]{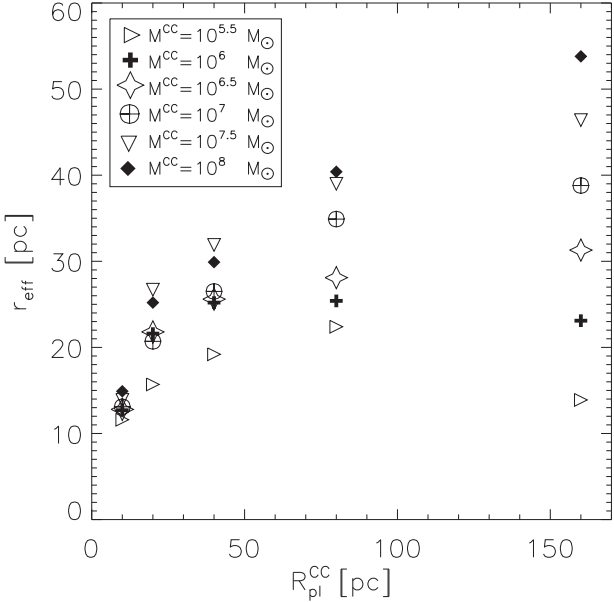}
\caption{Effective radius $r_{\rm eff}$ of the merger object vs. Plummer radius of the initial CC, 
$R_{\rm pl}^{\rm CC}$, for different CC-masses, $M^{\rm CC}$ for orbit 1 after 5 Gyr.}
\label{fig_r_eff_R_CC_pl.eps}
\end{figure}

\subsection{Correlation of structural parameters of merger object with CC parameter space} \label{merger_CC}

Naturally the number of merged star clusters has a substantial influence on the structural parameters 
of the merger object. The fraction of the merged mass is compared to the initial CC mass in 
Fig. \ref{fig_M_encl_M_CC.eps} for varying CC sizes. For compact models ($\alpha \ge 0.1$), where almost every 
star cluster merges and mass loss is small (less than 20\%), the final mass of the merger 
object is comparable to the initial CC mass.
In contrast, the enclosed mass of extended CCs strongly depends on the initial CC mass and size. 
For the most extended CC models the merger object masses lie between 25 and 80 \% of the initial CC mass.
The smaller the CC mass and the more extended the CC the larger is the influence of the tidal field. 
The CC experiences a larger mass-loss and the merging process gets suppressed.  

\begin{figure}
\centering
\includegraphics[width=8cm]{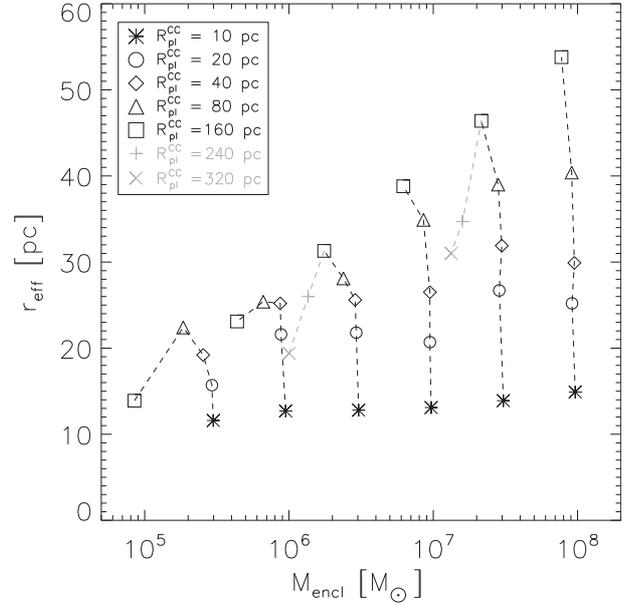}
\caption{Effective radii $r_{\rm eff}$ of the merger objects against the merger object masses $M_{\rm encl}$ 
for different CC sizes for orbit 1 after 5 Gyr. The dashed lines connect models with the same initial CC mass. 
Grey symbols represent additional models.}
\label{fig_comp_reff_mencl}
\end{figure}

Figure \ref{fig_r_eff_R_CC_pl.eps} shows the effective radius $r_{\rm eff}$ of the merger object vs. 
the Plummer radius of the CC, $R_{\rm pl}^{\rm CC}$, for different CC-masses, $M^{\rm CC}$. 
The effective radius corresponds to the projected half-mass radius, which is the radius within which  
half of the mass is included when projecting the merger object on the sky.
Compact CCs result in merger objects with effective radii comparable to the Plummer radius of the CC, 
while extended CCs result in merger objects with effective radii that are significantly smaller than 
the corresponding CC Plummer radius. A CC with a Plummer radius of $R_{\rm pl}^{\rm CC} = 10$ pc 
leads to merger objects with sizes between 10 pc and 15 pc while a CC with a Plummer radius of 
$R_{\rm pl}^{\rm CC} = 160$ pc yields an effective radius range of about 15 to 55 pc. The more 
extended the CC becomes the larger is the spread in the effective radii of the merger objects. 

For high CC masses of $M^{\rm CC} \ge 10^{6.5}$ M$_{\sun}$ the effective radii increase with increasing
Plummer radii. For the lower mass CCs, the effective radii decrease again for large Plummer radii 
$R_{\rm pl}^{\rm CC}$ = 160 pc.

\subsection{Trends in the $r_{\rm eff}$ vs. $M_{\rm encl}$ space}\label{reffvsmencl}

\begin{figure}
\centering
\includegraphics[width=8cm]{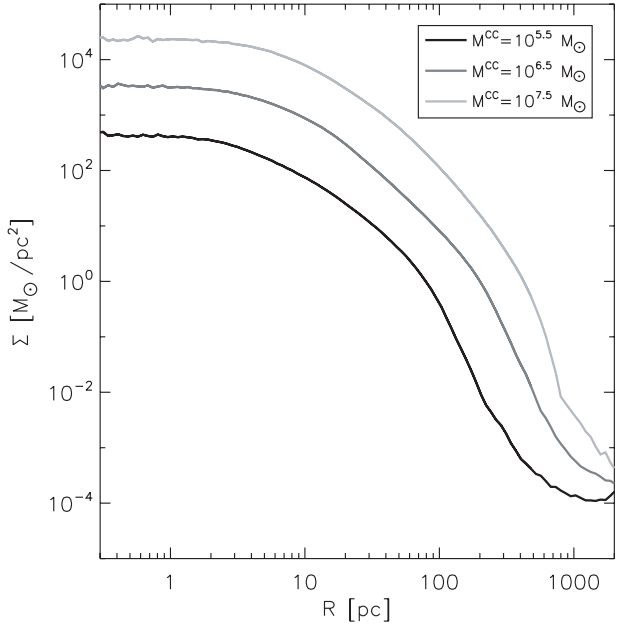}
\caption{Surface density profiles of the merger objects with $R_{\rm pl}^{\rm CC}$ = 80 pc and 
$M^{\rm CC} = 10^{5.5}, 10^{6.5}$, and $10^{7.5}$ M$_{\sun}$ for orbit 1 after 5 Gyr.}
\label{sb_profiles2}
\end{figure}

The parameter space of the CC models covers the $R_{\rm pl}^{\rm CC}$ vs. $M^{\rm CC}$ space
uniformly (Fig.~\ref{fig_grid}). The corresponding $r_{\rm eff}$ vs. $M_{\rm encl}$ space 
of the merger objects is shown in Fig.~\ref{fig_comp_reff_mencl}. 
For the most compact CC models with Plummer radii of $R_{\rm pl}^{\rm CC} = 10$ pc the effective radii 
and masses of the merger objects are very similar to the Plummer radii and masses of the CC, while 
for the most extended models $r_{\rm eff}$ strongly increases with increasing $M_{\rm encl}$.
Figure \ref{sb_profiles2} shows surface density profiles of models with 
$R_{\rm pl}^{\rm CC}$ = 80 pc and masses of $M^{\rm CC} = 10^{5.5}, 10^{6.5}$, and $10^{7.5}$ M$_{\sun}$. 
The surface density profiles are well represented by King profiles. The structural parameters 
central surface density, core radius, effective radius and tidal radius increase significantly 
with mass.

\begin{figure}
\centering
\includegraphics[width=8cm]{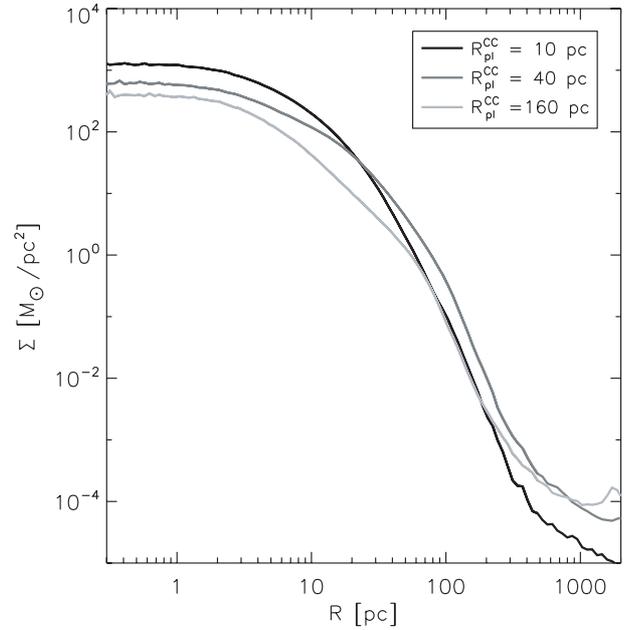}
\caption{Surface density profiles of the merger objects with $M^{\rm CC} = 10^{5.5}$ M$_{\sun}$ 
and $R_{\rm pl}^{\rm CC}$ = 10, 40, and 160 pc for orbit 1 after 5 Gyr.}
\label{sb_profiles}
\end{figure}

For a given CC mass, increasing the size of the CC results in a larger mass loss and larger effective 
radii of the resulting merger objects. For the lowest-mass models the effective radii decrease again
for the largest $R_{\rm pl}^{\rm CC}$. 
Figure \ref{sb_profiles} illustrates how the surface density profiles of the merger objects 
change with the CC Plummer radius ($R_{\rm pl}^{\rm CC}$ = 10, 40, and 160 pc) for a CC mass of
$M^{\rm CC} = 10^{5.5}$ M$_{\sun}$. The merger objects show King-like profiles. Increasing 
$R_{\rm pl}^{\rm CC}$ from 10 pc to 40 pc leads to a lower central surface density and larger 
values in the outer parts resulting in a larger effective radius. The merger object with 
$R_{\rm pl}^{\rm CC}$ = 160 pc suffered a major mass-loss, which leads to considerably lower 
surface densities especially at intermediate radii (5 to 50 pc) resulting in a lower effective 
radius.

For the very extended, low-mass models, the parameter $\beta$ is much larger than 
one, i.e. a considerable number of star clusters of the initial CC is located outside the tidal 
radius, leading to a rapidly decreasing number of merged star clusters. \cite{bruens10} studied 
such a turnover in detail for the Milky Way EC NGC\,2419 and found that the turnover occurs at 
those $R_{\rm pl}^{\rm CC}$, where the parameter $\beta$ is sufficiently large to allow entire 
star clusters to escape the merging process.

As high mass models have larger tidal radii, their size continuously increases with CC size up 
to $R_{\rm pl}^{\rm CC} = 160$ pc. However, increasing the CC size further will eventually result 
in decreasing $r_{\rm eff}$ also for high-mass models. This is demonstrated for two additional models 
for CC masses of $M^{\rm CC} = 10^{6.5}$ M$_{\sun}$ and $M^{\rm CC} = 10^{7.5}$ M$_{\sun}$. 
The CC sizes were extended to $R_{\rm pl}^{\rm CC} = 240$ pc and  $R_{\rm pl}^{\rm CC} = 320$ pc 
(grey symbols in Fig.~\ref{fig_comp_reff_mencl}). For both CC masses the results show a clear turnover 
in the effective radii of the merger objects for these large CC sizes.
 
The turnover leads to degenerate states in the merger-object space, as a relatively compact CC can
produce the same merger object as a more massive CC having a significantly larger CC size.

The turnover is a general feature of the merging scenario which occurs when a significant fraction of star 
clusters is located beyond the tidal radius of the initial CC. Therefore the merging star cluster scenario 
predicts for each CC mass an upper size limit of the merger objects. The exact CC sizes, where the turnover 
occurs, will also depend considerably on the initial configuration, i.e. the exact distribution of star 
clusters in the complex, the number of star clusters constituting the CC, and the orbit.

\begin{figure}
\centering\includegraphics[width=8.9cm]{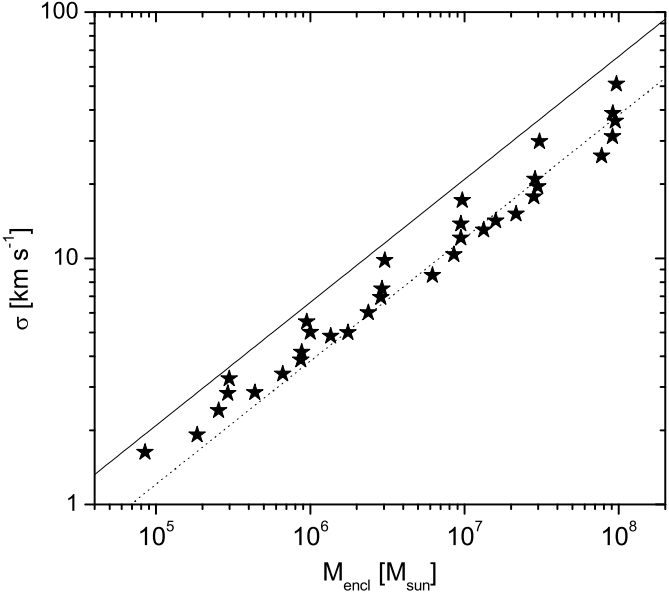}
\caption{Global line-of-sight velocity dispersion $\sigma$ of the merger objects as a 
function of the enclosed mass $M_{\rm encl}$ for orbit 1 after 5 Gyr. The solid and the dotted line show the 
scaling relation (Eq. \ref{Mdyn}) for objects with an effective radius of 10 pc and 30 pc, respectively.}
\label{sigmamodels}
\end{figure}

\subsection{Velocity dispersion and dynamical mass} \label{sigma}

The global line-of-sight velocity dispersion, $\sigma$, is an important observable
parameter as it can be used in combination with $r_{\rm eff}$ to estimate the dynamical mass, 
$M_{\rm dyn}$, of a star cluster. According to \citet{spitzer}, $M_{\rm dyn}$ can be estimated by
\begin{eqnarray}
M_\mathrm{dyn} \approx 9.75~\frac{r_\mathrm{eff}~\sigma^2}{G}\label{Mdyn},
\end{eqnarray}
where G is the Gravitational constant. Figure \ref{sigmamodels} shows the global line-of-sight 
velocity dispersion, $\sigma$, as a function of the enclosed mass of the merger objects. 
The turnover, which was discussed in the previous section, is clearly seen also in the velocity 
dispersion, as $r_{\rm eff}$ and $\sigma$ are not independent. For a given mass, an increasing $r_{\rm eff}$ 
results in a decreasing $\sigma$. The solid and the dotted line in Fig. \ref{sigmamodels} show the 
scaling relation of $\sigma$ versus mass for objects with an effective radius of 10 pc and 30 pc, 
respectively.

For our models, the dynamical masses calculated according to Eq. \ref{Mdyn} are
in very good agreement with the enclosed masses with a scatter of about five percent. 
The small deviations are due to slight deviations from virial equilibrium and due to the fact
that Eq. \ref{Mdyn} is only a rough estimate of the dynamical mass.

\begin{figure}
\centering
\includegraphics[width=8cm]{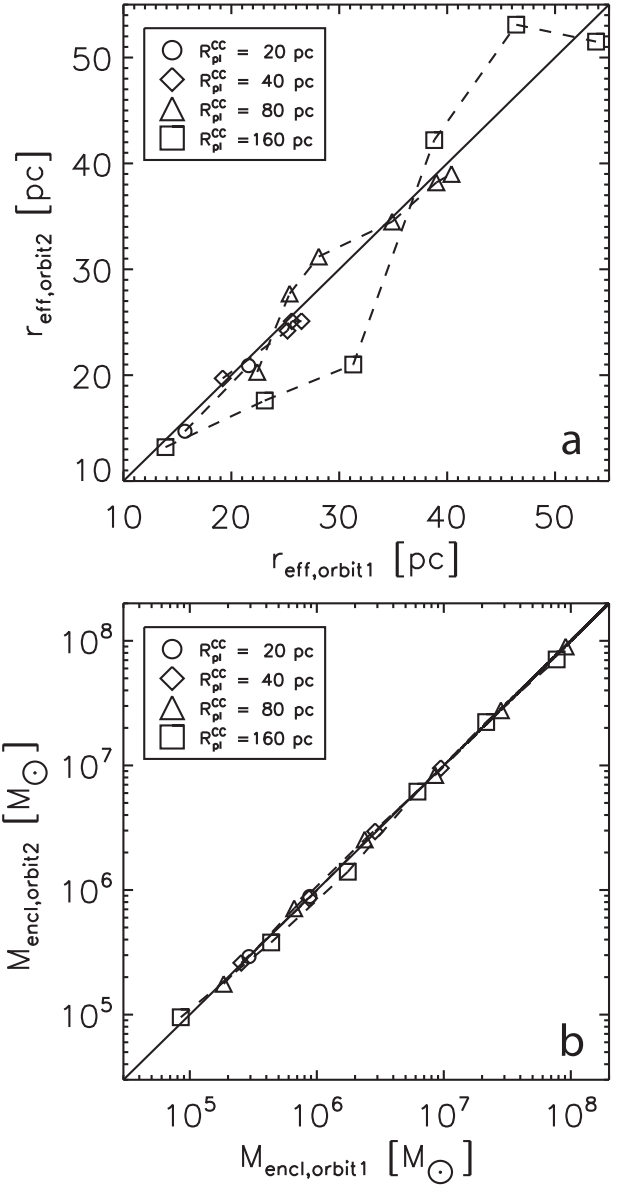}
\caption{\textbf{(a)} Effective radius $r_{\rm eff,orbit2}$ of the inclined orbit vs. effective radius $r_{\rm eff,orbit1}$
of the polar orbit for four CC sizes. \textbf{(b)} Enclosed mass $M_{\rm encl,orbit2}$ of the inclined orbit vs. enclosed mass 
$M_{\rm encl,orbit1}$ of the polar orbit for different CC sizes. For the solid line the values of orbit1
are equal to those of orbit2.}
\label{fig_result_comp_orbits}
\end{figure}

\subsection{Impact of polar orbit} \label{orbit2}

We recalculated 18 CC models on an inclined orbit (see Fig. \ref{fig_orbits}, Orbit 2) to estimate 
its impact on the structural parameters of the merger objects. In order to save computing time we 
only recalculated CC models where the inclination of the orbit is expected to have a measurable effect. 
These are the extended CC models with large CC-crossing-times, large values of $\beta$, and low
values of $\alpha$. These models are indicated by open circles in Fig. \ref{fig_grid}. 

Figure \ref{fig_result_comp_orbits}a compares the effective radii of the inclined orbit with
those of the polar orbit evaluated after 5 Gyr. Both orbits produce merger objects with 
comparable sizes. Only the most extended models ($R_{\rm pl}^{\rm CC} = 160$), which are most 
sensitive to the tidal field, show significant deviations between both orbits. 

Figure \ref{fig_result_comp_orbits}b compares the enclosed mass of the merger objects of the 
inclined orbit with those of the polar orbit. While there are some deviations of $M_{\rm encl}$ 
between the two orbits for the most extended models, the values for both orbits correlate very well.

The results of the inclined orbit as presented in Fig. \ref{fig_result_comp_orbits} demonstrate
that the inclination of the orbit has no significant influence on the overall results and trends 
of our parametric study.

\begin{figure}
\centering
\includegraphics[width=8.9cm]{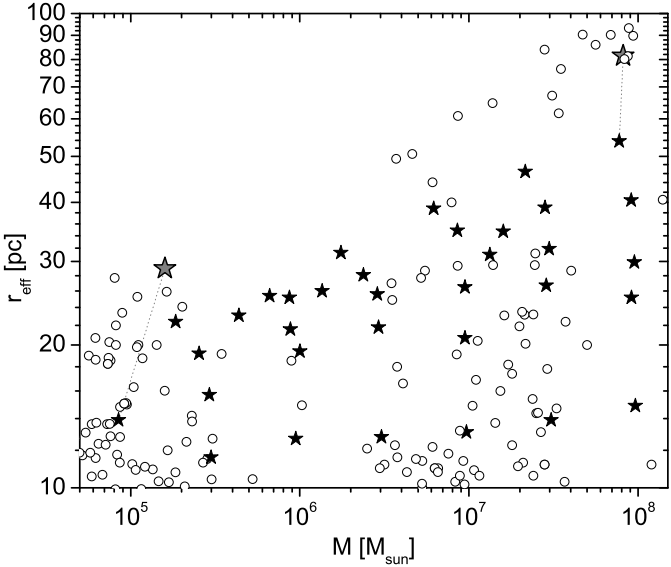}
\caption{The diagram shows $r_{\rm eff}$ as a function of mass of the observed ECs and UCDs (circles) 
and the modeled merger objects (stars), combining Figs. \ref{fig_ucdgc} and \ref{fig_comp_reff_mencl}.
Two additional models with $R_{\rm pl}^{\rm CC} = 160$ calculated on a circular orbit at 60 kpc 
($M^{\rm CC} = 10^{5.5}$ and $10^{8.0}$ M$_{\sun}$ ) are plotted as grey stars. The dotted lines 
connect the additional models on a circular orbit with the corresponding models on an eccentric orbit.}
\label{simsandobs}
\end{figure}

\section{Discussion and conclusions} \label{discussion}

We systematically scanned a suitable parameter space for CCs and investigated their future 
evolution. The varied sizes and masses of the CCs covered a matrix of 5x6 values with 
CC Plummer radii between 10 - 160 pc and CC masses between $10^{5.5}$ - $10^{8}$ M$_{\sun}$. 
The results presented in Sect. \ref{results} demonstrate that all simulations end up with  
stable merger objects, which show a general trend of increasing effective radii with increasing 
mass. Despite the large range of input Plummer radii of the CCs the effective radii of the merger 
objects are constrained to values between 10 and 20 pc at the low mass end and to values 
between 15 and 55 pc at the high mass end. 
The turnover in the $r_{\rm eff}$ vs. $M_{\rm encl}$ space (see Fig. \ref{fig_comp_reff_mencl}) 
depends on the mass of the initial CC and occurs at larger sizes for higher masses. 
The turnover leads to a degeneracy in the $r_{\rm eff}$ vs. $M_{\rm encl}$ space of the merger 
objects, i.e. very different CC parameters can result in a comparable final 
merger object. In addition the turnover leads to a higher probability for merger objects to 
have intermediate effective radii.

Figure \ref{simsandobs} shows $r_{\rm eff}$ as a function of mass of the observed ECs and UCDs (circles)
and our models (stars), combining Figs.~\ref{fig_ucdgc} and \ref{fig_comp_reff_mencl}. 
The vast majority of the observed ECs and UCDs are located within the parameter space covered by the 
modeled merger objects. Only the very extended objects at $M_{\rm EC} \approx 10^5$ M$_{\sun}$
and the extremely extended UCDs between $M_{\rm UCD} = 10^7$ and $10^8$ M$_{\sun}$ are outside 
the parameter space covered by this study. 

In order to verify that less eccentric orbits would produce more extended objects, we calculated 
two additional models. For the most extended models with the lowest and the highest mass 
($R_{\rm pl}^{\rm CC} =$ 160 pc, $M^{\rm CC} = 10^{5.5}$ and $10^{8}$ M$_{\sun}$) we calculated the 
evolution on a circular orbit at a galactocentric distance of 60 kpc. The corresponding merger objects 
have considerably larger effective radii of 29 pc and 82 pc (see Fig. \ref{simsandobs})
than the merger objects on the eccentric orbit, which have effective radii of 14 and 54 pc. 
Due to the lower gravitational field, the masses of the merger objects on the circular orbits are larger 
than those of the eccentric orbits. For the lowest mass model the enclosed mass increases from 27\% to
50\%. These results demonstrate that very extended objects like the M31 ECs found by \cite{huxor04} 
and the very extended, high-mass UCDs can be explained by merged cluster complexes in regions with 
low gravitational fields at large galactocentric radii. 

\begin{figure}
\centering
\includegraphics[width=8.9cm]{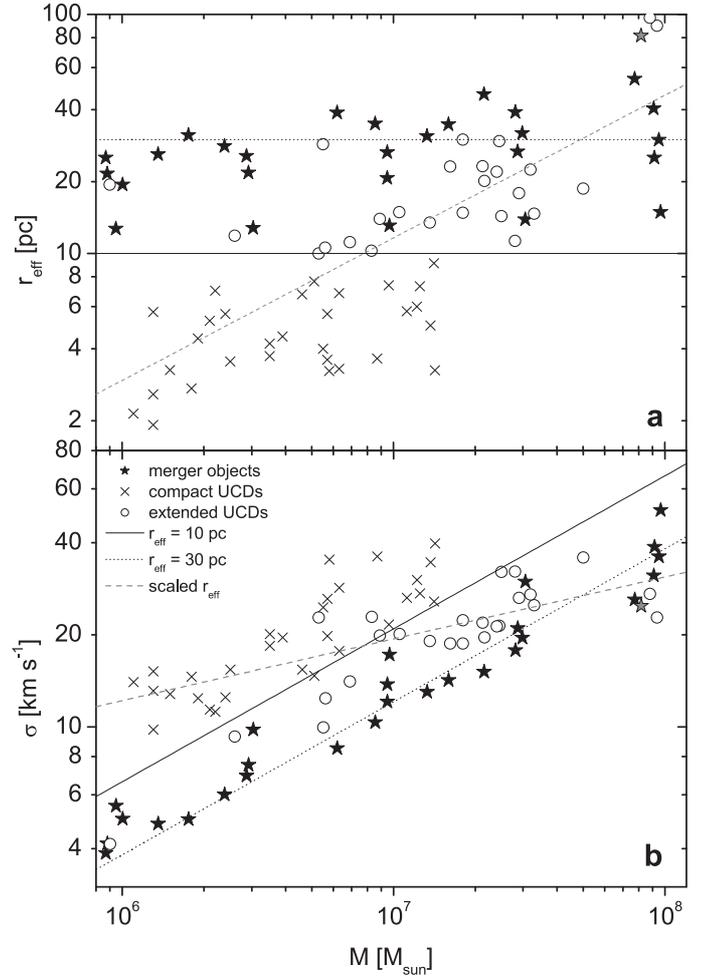}
\caption{\textbf{(a)}: Effective radii as a function of mass of the merger objects (stars) and of
UCDs with observed velocity dispersions (crosses for compact UCDs with $r_{\rm eff} <$ 10 pc, 
circles for extended UCDs and NGC\,2419). \textbf{(b)}: Global line-of-sight velocity dispersion $\sigma$ 
as a function of the mass of the same objects as in the above figure. The solid and the dotted line 
show the scaling relation for objects with an effective radius of 10 pc and 30 pc, respectively (see Eq. \ref{Mdyn}). 
The dashed grey curve indicates the general trend of $\sigma$ for an increasing $r_{\rm eff}$  
\cite[as parameterized by][see Fig. \ref{fig_ucdgc}]{dabringhausen08}.}
\label{sigmaandobs}
\end{figure}

The evolution of ECs in a weak gravitational environment has also been studied by \cite{hurley}, 
who performed direct N-body models of extended low-mass star clusters incorporating a stellar mass 
function and stellar evolution. They concluded that extended star clusters with an initial mass 
of 5.8 $10^{4}$ M$_{\sun}$ are sufficiently stable to survive a Hubble-time in a weak gravitational field 
environment. \cite{bekki04} modeled the first 70 Myr of the merging of high-mass star clusters without 
an external tidal field. Their finding of triaxial shapes of UCDs is most likely related to the relatively
short simulation time. Figure \ref{fig_time_evol} illustrates that the merger objects reach a spherically
symmetric shape only after a few Gyr of evolution. \cite{bekki04} find a general trend of increasing 
$r_{\rm eff}$ and velocity dispersions, $\sigma$, with increasing mass. Their effective radii increase from
about 8 pc at a mass of 4 $10^{6}$ M$_{\sun}$ to values of the order of 20 pc at masses of 4 $10^{7}$ M$_{\sun}$.

Figure \ref{sigmaandobs} shows only those UCDs and NGC\,2419 where observations of 
the effective radius and the global line-of-sight velocity dispersion are available 
\citep{hasegan,evstigneeva07,mieske08,baumgardt,hau,taylor}. 
Figure \ref{sigmaandobs}a, which shows $r_{\rm eff}$ versus mass, demonstrates that most of 
the very extended UCDs with $r_{\rm eff} >$ 20 pc shown in Figure \ref{simsandobs} have so far no 
observed velocity dispersions. 
Figure \ref{sigmaandobs}b shows observed global line-of-sight velocity dispersions of NGC\,2419 and UCDs 
and of our merger objects (see Sect. \ref{sigma}) as a function of mass.
The models have a steeper relation of $\sigma$ vs. mass than the observed UCDs. This is due to the fact that
we modeled solely extended objects with $r_{\rm eff} >$ 10 pc, while UCDs show a general trend of 
increasing $r_{\rm eff}$ with increasing mass (see Fig. \ref{fig_ucdgc}). The mean effective radii of 
the UCDs considered in Figure \ref{sigmaandobs} increase from about 5~pc for masses in the interval
$10^{6.0}$ to $10^{6.5}$ M$_{\sun}$ to about 16~pc for masses between $10^{7.0}$ and $10^{7.5}$ M$_{\sun}$.
The grey dashed line in Figure \ref{sigmaandobs}, which combines Eq. \ref{Mdyn} with the parameterization 
of $r_{\rm eff}$ vs. mass from \cite{dabringhausen08}, is a good representation of the $\sigma$ vs. mass 
relation of the observed UCDs. The results of \cite{bekki04} are closer to the observed values as their merger
objects have much smaller effective radii than our objects. However, it should be kept in mind that 
a considerable amount of UCDs with large sizes ($r_{\rm eff} >$ 20 pc) do not have observed velocity 
dispersions, yet.

The continuous distribution of CC masses used in our parametric study results in a continuous 
distribution of masses of merger objects. In contrast, the observed masses of ECs and UCDs show 
clear accumulations near masses $10^5$ M$_{\sun}$ and between $10^7$ and $10^8$ M$_{\sun}$ and a 
very low number of ECs near $10^6$ M$_{\sun}$ (see Fig. \ref{simsandobs}). 
A straightforward interpretation on the basis of the proposed formation scenario would suggest 
that the mass distribution of ECs and UCDs facilitates direct conclusions on the mass spectrum 
of the CCs, which produced the ECs and UCDs. 

An interpretation of the available data on ECs and UCDs must be done, however, with great care, 
as the underlying datasets (see Sect. \ref{observations}) are highly inhomogeneous and incomplete.

Due to the limited field of view of the Hubble Space Telescope, most extragalactic studies on 
GCs and ECs cover only (a part of) the optical disk of the respective galaxies.
The ECs discussed in this paper are, however, halo objects located far from the optical disk of the 
galaxies. In the Milky Way, 9 out of 13 ECs have galactocentric distances greater than 20 kpc 
\citep{harris}. The only massive EC ($M_{\rm EC} \approx 10^6$ M$_{\sun}$ and 
$r_{\rm eff} \approx 20$ pc) of the Milky Way, NGC2419, is located at a distance of about 92 kpc. 
A similar trend has been shown for the other two disk galaxies in the Local Group: 12 out of 
13 ECs associated with M31 and both ECs found in M33 have projected distances 
well outside the optical disks of these galaxies \citep{huxor08,stonkute,huxor09}. While halo ECs 
might be found by chance in projection to the main body of a galaxy, the probability is relatively low: 
if a survey covers a projected area of 20 kpc by 20 kpc and a line-of-sight of $\pm$100 kpc is 
considered, the resulting volume, which is covered by the survey, is only about two percent of the 
volume of a sphere of radius 100 kpc, wherein the ECs would be distributed. 

In addition, \cite{larbro00} and \cite{larsen02} have discovered a population of ECs co-rotating with 
the disk of the lenticular galaxy NGC1023. These so-called faint fuzzies have similar structural parameters
as halo ECs and are therefore not easily distinguishable from halo ECs projected onto the disk on the basis 
of imaging data alone. A fair fraction of ECs found in extragalactic surveys might therefore be associated 
with the disks and not the halos of these galaxies.\cite{burkert} analyzed the kinematics of 
the faint fuzzies and concluded that they form a ring-like structure within the galactic disk of NGC\,1023 
and that this ring was probably formed during a galaxy-galaxy interaction comparable to the Cartwheel galaxy. 
A detailed discussion of faint fuzzies in the context of merged CCs is given in \cite{bruens09},
who demonstrated that the observed structural parameters of the faint fuzzies are in excellent
agreement with the merged CC scenario.

Another reason for incompleteness is the difficulty of distinguishing ECs from background 
galaxies. The GC surveys covering 100 galaxies of the Virgo Cluster \citep{jordan05} and 43 galaxies 
of the Fornax Cluster \citep{masters10} applied a size limit of $r_{\rm eff} <$ 10 pc to reduce the 
contamination of background galaxies. Thereby, they excluded also all ECs from their GC catalogs. 
\cite{peng} used the same Virgo Cluster survey data as \cite{jordan05} to search for diffuse star 
clusters and found e.g. in the galaxy \object{VCC798} about 30 ECs, where \cite{jordan05} found 211 
compact GCs. \cite{peng} have demonstrated that hundreds of ECs await detection in galaxy clusters. 
Without follow-on spectroscopy to determine radial velocities, it cannot be decided whether 
these ECs are associated with the main body or the halo of the galaxies. 

While lower-mass ECs up to $M_{\rm EC} \approx 10^6$ M$_{\sun}$ are often overlooked in surveys,
much effort has been made to detect and to analyze stellar objects of $M \approx 10^7$ M$_{\sun}$
since the discovery of UCDs in the Fornax Cluster by \cite{hilker99} and \cite{drinkwater00}. 
Therefore, the incompleteness for UCDs is expected to be considerably lower.

In conclusion, the rapidly increasing number of detected ECs and UCDs associated with various 
types of galaxies in different environments offers a new perspective to the process of cluster                                                                           
formation and galaxy evolution. Since galaxy-galaxy mergers are anticipated to have been more 
common during early cosmological times it is expected that star-formation in cluster complexes 
has been a significant star-formation mode during this epoch. Our work provides a unification 
of the compact state of young star clusters and the extended size of ECs by allowing ECs to 
form from a CC of compact star clusters.
If the formation scenario suggested in this paper is correct, a statistical analysis of the observed 
ECs und UCDs with respect to the results of the models has the potential to shed light on the 
mass spectrum of the initial CCs and thereby on the interaction history of galaxies at this 
cosmologically important epoch. 

Considerably larger, more homogeneous, and more complete datasets of ECs and UCDs and more detailed
observations of CCs are, however, necessary to draw statistically significant conclusions on their 
origin.

\begin{acknowledgements}
We thank the anonymous referee for his helpful comments, which lead to a considerably improved paper.
 The work of this paper was supported by DFG Grants KR\,1635/14-1 and KR\,1635/29-1.  
\end{acknowledgements}

\end{document}